\documentstyle[preprint,aps,epsf]{revtex}

\def\by#1#2{{\displaystyle {#1}\over \displaystyle {#2}}}

\preprint {IMSc/2002/01/01}
\begin{document}
%\draft
\title{Implications of recent solar neutrino observations: an analysis
of charged current data}

\author{C. V. K. Baba\footnote{Permanant address: Nuclear Science 
Centre, New Delhi 110 067}, D. Indumathi and M. V. N. Murthy}

\address
{The Institute of Mathematical Sciences, Chennai 600 113, India.\\
}
\date{Jan 1, 2002}
\maketitle
\begin{abstract}

We have analysed the recent results from the observation of charged
current $\nu_e d \rightarrow e^{-} p p$ events from solar neutrinos
by the Sudbury Neutrino Observatory SNO assuming neutrino oscillations
with three active flavours. The data seem to prefer a low mass-squared
difference and large mixing angle solution (the so-called LOW solution)
in (12) parameter space. However, when combined with the Gallium charged
current interaction data from Gallex and GNO, distinct ($1\sigma$)
allowed regions corresponding to the large mixing angle (LMA) and small
mixing angle (SMA) appear while the LOW solution is disfavoured upto
$3\sigma$ standard deviation. The physical electron neutrino survival
probability corresponding to these best fit solutions are then determined
and analysed for their energy dependence.

\end{abstract}

\pacs{PACS numbers: 14.60.Pq, 26.65.+t}

%\narrowtext

\section{Introduction}

The recent announcement of results on the observation of charged current
(CC) neutrino events from the sun by the Sudbury Neutrino Observatory
(SNO) \cite{sno} has lead to a flurry of activity in understanding the
implications of the data on neutrino oscillations. Several authors have
concentrated on the global analysis \cite{analysis} of the solar neutrino
data from all existing observations and the refinement in the allowed
oscillation parameter space. Typically, such analyses have refined the
already existing allowed regions in parameter space---the so-called
small mixing angle (SMA), large mixing angle (LMA), Low Mass (LOW) and
the `just so'  solutions. However, not as much attention has been paid
to how individual experimental results perform with respect to the best
possible global fits.

This paper is not devoted to providing yet another global best fit. We
concentrate rather on the {\it physical} quantity of interest, viz.,
the neutrino survival probability. Since we restrict ourselves to
the CC data, we focus on the electron neutrino survival probability,
$P_{ee}(E_\nu)$. The question that we seek to address is, what are the
constraints from data on the energy dependence of $P_{ee}$. In order
to answer this question, we use the energy integrated CC data from
the combined Gallium experiments of GALLEX and GNO\footnote{We have used
the cited average of these data; the SAGE Gallium experiment \cite{sage}
is also consistent with these data.} \cite{gg} (henceforth
referred to as the Ga experiments) as well as the recent CC observations
from SNO \cite{sno}. We also briefly discuss the constraints arising
from the Homestake Chlorine experiment \cite{homestake} (henceforth called
Cl). We do not include the results from Super-Kamiokande in our analysis
since the observed elastic scattering process has both CC and
neutral current (NC) components. We use the standard solar model of BP2001
in our theoretical analysis; the flux data are available in tables at
the web-site \cite{bp2001}.

Of these experiments, the Ga experiments observe
neutrinos over the widest range of energies. However, the Ga data
contain no energy information. Theoretically, the bulk of
the events may be attributed to electron neutrinos with energy less than
1 MeV with less than 10\% coming from neutrinos with energy more than
5 MeV. On the other hand SNO observes only the ${}^8$B neutrinos above a
threshold of about 8 MeV. It is also sensitive (indirectly) to the
neutrino energy. In short, these experiments are sensitive to different
source reactions in the sun. We can therefore get good information by
analysing the combined data as well as the individual data sets. The Cl
experiment is sensitive to the (second) ${}^7$Be line apart from the B
neutrino flux. Like Ga, it does not contain any energy
information. However, it turns out that Cl and SNO yield similar
likelihood curves in the allowed parameter space.

We first analyse the experiments individually and then the combined Ga
and SNO data. We have used simple expressions for the likelihood
contours where the $\chi^2$ is defined using diagonal errors only; all
correlated errors are ignored. This approximation is obviously valid for
SNO since there is exactly one contributing source flux; it remains
true to good accuracy for Ga since the dominant low energy pp source
flux has very small errors. Finally, since there is only 10\% overlap
between the Ga and SNO source fluxes, the neglect of correlated errors
is valid while analysing the combined data as well. This is not true for
the combined Ga, Cl and SNO data sets; hence we have not presented the
results of such a combined analysis here.

Assuming that neutrino oscillations are the source of the discrepancy
between data and theory, we find that the likelihood contours in the
allowed parameter space for the Ga experiments differ substantially from
the SNO and Cl ones. Consequently, it turns out that the electron neutrino
survival probability (the CC analysis depends only on this probability)
corresponding to the best fit parameters ($\chi^2$ minimum) has a very
different energy dependence for the fits coming from analysing the Ga
or SNO data alone and the combined fits. This is true especially for the
so-called LMA solution where the SNO prefers a rather flat $P_{ee}$ while
Ga shows a distinct energy dependence.  This leads to the question whether
the low energy $pp$ neutrinos behave very differently from the higher
energy ones. If so, it is possible that a measurement of the (second)
${}^7$Be line spectrum will be able to settle this issue. For example,
an almost vanishing ${}^7$Be flux ($< 10$--20\% of the expected flux)
will point unambiguously to the SMA as the preferred solution to the
neutrino oscillation problem. A similar analysis for the best-fit global
solutions may be found in Ref.~\cite{BL}.

In Section 2, we present the relevant formulae for the event rate as
well as for the survival probability of (e-flavour) neutrinos in the
3-flavour oscillation scheme. In Section 3 we present the
results of the numerical calculation using the standard solar model of
BP2001 \cite{bp2001} for the solar neutrino fluxes. Section 4 contains
the summary and discussion. 

\section{Relevant Formulae}

We briefly list the relevant formulae in a 3-flavour oscillation
scenario to explain the discrepancy between data and theory. The mixing
is expressed as,
$$
\left(\begin{array}{lll} \nu_e & \nu_\mu & \nu_\tau \\ \end{array}
\right)^{\rm T} = V \times
\left(\begin{array}{lll} \nu_1 & \nu_2 & \nu_3 \\ \end{array}
\right)^{\rm T}~,
$$
where the mass eigenstates $\nu_i$ have masses $m_i$ and the mixing
matrix is parametrised in terms of three angles\footnote{We ignore the
CP violating phase here.} as a set of 3 independent $2\times 2$ rotations:
$$
V = V_{23} (\psi) \times V_{13} (\phi) \times V_{12} (\omega)~.
$$
The angle $\psi$ does not occur in expressions for solar neutrinos since
only $\nu_e$ neutrinos are produced in the sun. 

Since we are interested only in the CC events, the only relevant
probability is the electron neutrino survival probability, $P_{ee}$. This
can be expressed in terms of the (12) and (13) mixing angles, $\omega$
and $\phi$, as well as the (12) mass squared difference, $m_2^2 - m_1^2 =
\delta_{12}$. We have
\begin{equation}
P_{ee} = \sin^2\phi \sin^2 \phi_m + \cos^2\phi \cos^2 \phi_m \left[
\by{1}{2}
+ \left(\by{1}{2} - P_{LZ}\right) \cos 2 \omega \cos 2 \omega_m \right]~.
\label{eq:pee}
\end{equation}
Here the subscript $m$ includes the usual (solar) matter effects. We
do not include earth matter effects here; this affects only the
`LOW' solution and that too, at a small level. Here $P_{LZ}$ is
the non-adiabatic jump probability \cite{KuoP} and is relevant near
resonance. While $\cos2\phi_m \sim \cos2\phi$, $\cos2\omega_m$ depends
on the vacuum mixing angles as well as $\delta_{12}$ and the matter term
$A(E_\nu)$. Hence it is energy dependent. Note that we have used the
phase averaged expression for $P_{ee}$; this is allowed as long as we
restrict ourselves to $\delta_{12} > 10^{-8}$ eV${}^2$. Hence we do not
consider the `just-so' solutions in this paper. Finally, $A(E_\nu)$ is also
density dependent \cite{KuoP}. The solar density varies with distance;
however, we have used average values of density corresponding to the peak
production of different source fluxes since we have averaged over the
production region. The results are not sensitive to this approximation
\cite{KuoP}.

The survival probability appears in the expression for the ratio:
\begin{equation}
R^{\rm th} = \by{\int {\rm d}E_\nu \, P_{ee}(E_\nu) \, \phi(E_\nu)
\, \sigma(E_\nu)}{\int {\rm d}E_\nu \, \phi(E_\nu) \, \sigma(E_\nu) }~,
\end{equation}
where the factor $(P_{ee} \, \phi)$ in the numerator represents the depleted
flux due to mixing and depends on the mixing parameters as shown; $\sigma$
is the cross-section for the relevant process in the detector; $\sigma$
may also be differential with respect to one or more variables as in
the case of the SNO data where we use ${\rm d}\sigma/{\rm d}E_e$, with
$E_e$ the scattered electron energy. There is a further smearing out
of this data due to the resolution function since the measured electron
energy $E_e$ is a gaussian distributed around the true electron energy
$E'_e$ \cite{sno}. The Ga cross-section is taken from tables available
in Ref~\cite{bahcallga}. The deuterium cross-section is taken from the
tables available in Ref.~\cite{Kubodera}.

The theoretical value of $R^{\rm th}$ calculated this way is then
compared with the ratio $R^{\rm exp}$ measured by the different
experiments\footnote{The SNO ratio is calculated using the same BP2001
model. The experimental event rate in SNUs for the Ga (actually the
average from Gallex and GNO) and Cl experiments were used and converted
to a ratio using the BP2001 calculated prediction.}. To avoid double
counting, the theoretical errors were factored into $R^{\rm th}$ while
the experimental numbers included only the statistical errors.

The significance of SNO and Gallex for mass and mixing parameters is 
obtained by separately determining the minimum $\chi^2$, where we use 
the standard  definition
\begin{equation}
\chi^2 = \sum_{\alpha} \frac{(R_{\alpha}^{\rm th} 
-R_{\alpha}^{\rm exp})}{\sigma_{\alpha}^2}~,
\end{equation}
where the explicit parameter dependence of the ratio is $R_{\alpha}^{\rm
th}= R_{\alpha}^{\rm th}(\delta_{ij},\theta_{ij})$ for a given set
of mass squared differences, $\delta_{ij}$, and mixing angles,
$\theta_{ij}$. The range of the subscripts is determined by the
number of neutrino flavours. For $P_{ee}$ we have the three parameters
$\delta_{12}$, $\omega$ and $\phi$. The subscript $\alpha$ runs over the
set of data points, $R_{\alpha}^{exp}$, of a given experiment(s). The
theoretical and experimental errors are added in quadrature in
$\sigma_{\alpha}^2$. Likelihood contours are drawn using
$$
{\cal{L}} = \exp\left[ -\chi^2/2 \right] ~.
$$
While ${\cal{L}}$ by itself has no meaning (it is the log likelihood or
$\chi^2$ itself that is relevant), $n\sigma$ likelihood contours can be
drawn corresponding to $\chi^2 = \chi^2_{\rm min} + n^2$. These contours
have the usual probability interpretation of allowed regions.  We use
the parameters corresponding to the best fit (and within $1\sigma$ of the
best fit) to find the preferred survival probability.

As we have observed, all charged current events are directly proportional
to the electron neutrino survival probability. This is not true for
the case of the Super-Kamiokande data in the presence of an additional
sterile flavour; analysis of this data would require an additional
unknown, the oscillation of $\nu_e$ into the sterile flavour. This is
why an analysis of charged current data provides the most restrictive
bounds on the mixing.

\section{Numerical Analysis and Results}

\subsection{Choice of data}

We have individually analysed the following sets of solar neutrino data:
CC data from SNO (referred to as SNO data), from the Gallium detectors
GALLEX and GNO (referred to as Ga data), and from the Chlorine detector
at Homestake (referred to as Chlorine data). All these experiments measure
charged current data (we have not analysed the SNO elastic scattering data
here) and hence are particularly tuned to the electron-type neutrinos.
The data are expressed as ratios of observed to expected event rates. We
use the BP2001 \cite{bp2001} standard solar model for the theoretical
calculation.

Energy integrated data from Ga and Cl (with $E_{\nu,{\rm min}} = 0.24$
and 0.814 MeV respectively) yield one data point from each
experiment while SNO (apart from an energy integrated value for the
ratio of observed to expected rate) also gives data in 11 bins for the
recoil electron energy starting from $E_e$ of about 7.25 MeV, with
bin size about 0.51 MeV (except for the last bin) \cite{sno}. The $Q$
value for the reaction $\nu_e \, d \to e p p$ is 0.93 MeV, so the parent
neutrino energy is at least 8 MeV; it is however useful to note that
the cross-section for this reaction peaks typically at $E_e = E_\nu
- 1.4$ MeV \cite{Kubodera} although the peak value is mildly energy
dependent. Hence there is good energy information in the SNO data.

As pointed out earlier, SNO sees only the ${}^8$B neutrinos from the sun,
Chlorine also sees most (about 90\%) of the Be neutrinos and Ga sees
both Be and B and much of the pp neutrinos. In our analysis, we have
ignored errors due to correlations between the different fluxes (that
would not affect SNO) contributing to the event rate at any given
experiment. We have also used only the statistical errors in the data
(since, typically, systematic errors are also correlated between data
bins). Hence the $\chi^2$ minima that we compute are a lower bound on
that which would have been obtained if the correlated errors were taken
into account.

Also, as pointed out in the Introduction, it is possible to do a combined
analysis of the Ga and SNO data while still ignoring correlated errors.
This has also been done.

\subsection{Allowed Parameter Space}

We begin by finding the allowed parameter space for the individual
data sets.  These are shown in Figs.~\ref{fig:1sigma} where the allowed
1$\sigma$ regions of $\delta_{12}$ and $\omega$ around the global minimum
$\chi^2$ are shown for $\phi = 0$. There is not much sensitivity to $\phi$
in the range allowed by {\sc chooz}, $\phi \le 9^\circ$ \cite{chooz}. The
2$\sigma$ regions (not shown in the figure) are quite large, especially
for the Ga and Cl data sets since there is exactly one data point
(total rate) available. While it may appear that one data point is being
fitted with two (or even three) free parameters for Ga and Cl, the
resulting restriction in parameter space occurs because of the
non-trivial energy dependence of $P_{ee}$ and the fact that different
source fluxes contribute differently at different energies. For the
SNO data, we have shown the region of parameter space allowed by taking
into account either all the 11 electron energy bins or the integrated
rate. Obviously, the allowed region is larger if only the average
(integrated) rate is taken into account, as shown in the figure. However
there is a good overlap between these two fits, indicating the stability
of the fits.

There is almost complete overlap between the allowed parameter
space from Chlorine and SNO data especially at large angles. The
contribution to Cl from the region below the SNO threshold comes roughly
equally from the (second) ${}^7$Be line spectrum and from a part of
the ${}^8$B flux. The overlap between the two sets therefore seems to
indicate that the nature of suppression in the SNO energy regime is
similar to that at lower energies, down to the Cl threshold. The SNO
energy-dependent data are more restrictive; the effect of this is to
allow a narrower strip of parameters than the Chlorine data.

What is interesting is that the preferred regions for Ga and Chlorine/SNO
are totally different as can be seen by comparing the relevant panels
of Fig.~\ref{fig:1sigma}. The so-called LMA solution for Ga has a
well-defined $\omega \sim 25^\circ$ but accomodates a large range
of $\delta_{12}$, $\delta_{12} > 10^{-5}$ eV${}^2$. Solutions
corresponding to the SMA exist although the SMA region is not distinct
from the LMA one; however, the LOW solution (corresponding to low mass
squared, large angle) is conspicuously absent in Ga.

Even more surprisingly, while SNO shows allowed regions in the regions
corresponding to SMA, LMA and LOW, these regions are connected with no
well-defined LMA or SMA minima. The global minimum for the entire region
is in fact the so-called LOW solution centred around $\delta_{12} \sim
6.7 \times 10^{-8}$ eV${}^2$ and $\omega \sim 29^\circ$.

In general, the SMA and LMA allowed regions of SNO and Ga are different.
Hence, when the Ga and SNO data are combined distinct regions in LMA
and SMA parts of the parameter space appear (see Table~1). Since Ga does
not have a LOW solution, the `LOW' solution (that is in fact the global
minimum of the fits to the SNO data) is completely disallowed by the
combination of Ga and SNO data by about 4$\sigma$.  The SMA solution
is marginally preferred over the LMA.  However, in our simple analysis
we have neglected correlations as well as earth matter effects. Though
small, this precludes us from preferring SMA over LMA; within the errors
of the analysis both are equally likely.

\begin{figure}[htpb]
\vskip 15truecm
\includegraphics{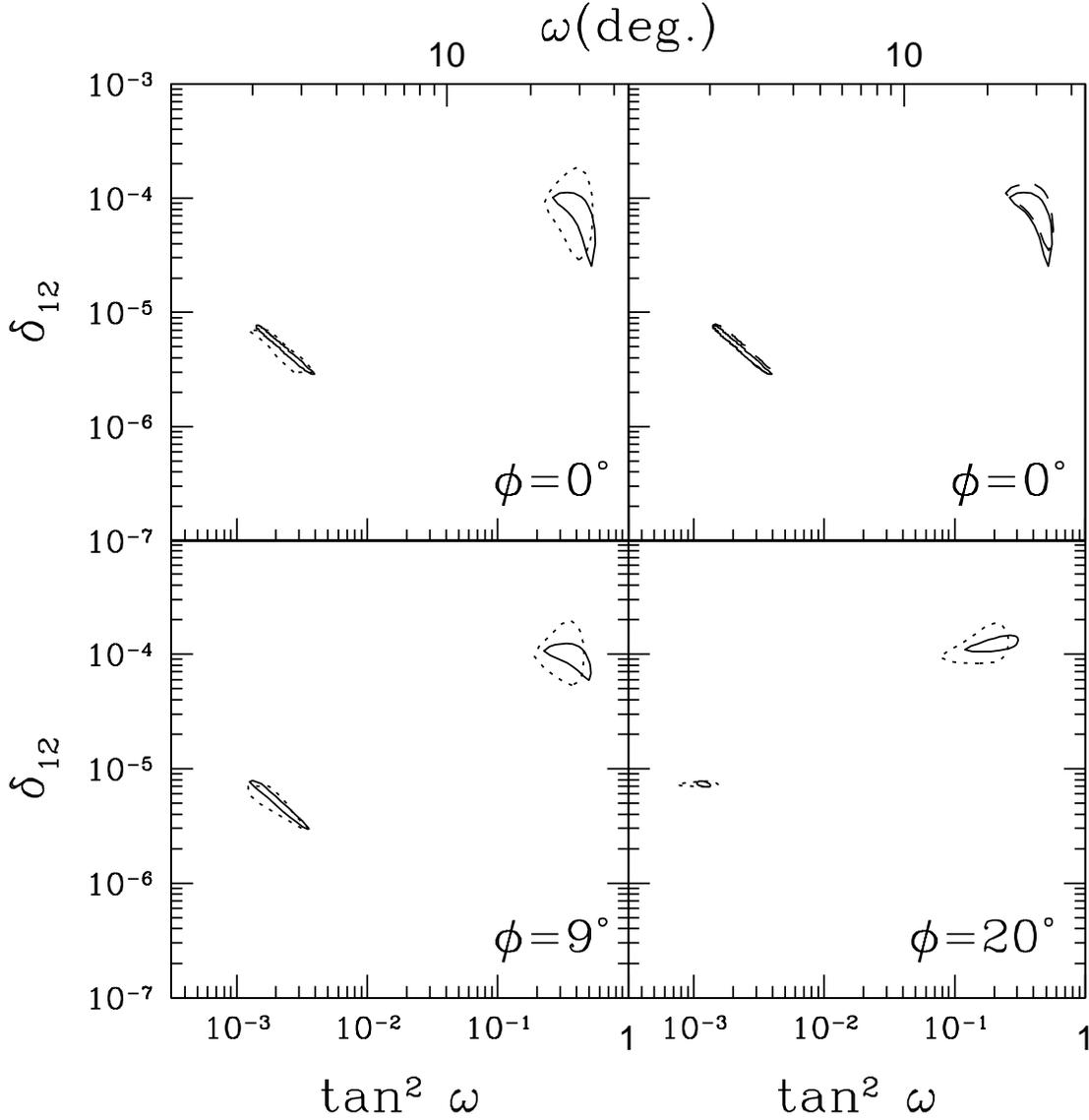}
	  \caption{Allowed 1$\sigma$ regions in $\delta_{12}$ and the
	  (12) mixing angle $\omega$ (both $\omega$ and $\tan^2\omega$
	  are shown) parameter space for the combined Gallium and SNO data
	  sets. The solid lines use the energy-binned SNO data while the
	  dotted lines use the integrated SNO data. Allowed regions are
	  shown for fixed values of $\phi = 0, 9, 20$ degrees. It is seen
	  that the allowed region shrinks with increasing $\phi$. The
	  top right panel shows the fits (at $\phi = 0^\circ$) obtained
	  by using all the SNO binned data (solid lines) and that from
	  excluding the first four bins where correlated errors are large
	  (dashed lines).}
\label{fig:lma} % fig2
\end{figure}

The result from the combined fit is shown in Fig.~\ref{fig:lma}. Note
that since we have ignored correlations between the fluxes, the allowed
regions are just the overlaps of the separately allowed regions. In the
figure, the 1$\sigma$ allowed contours (with respect to the local
$\chi^2$ minima) are shown in the $\delta_{12}$-$\omega$ plane
for fixed $\phi = 0^\circ$, $9^\circ$, and $20^\circ$. The solid
lines correspond to the combined analysis of Ga and SNO energy-binned
data, while the dotted lines correspond to the combined Ga and integrated
SNO data. As expected, the allowed region shrinks when the energy
dependence of the SNO data is taken into account. We have also analysed
the Ga data together with data in the higher energy bins of SNO (from
the 5th to the 11th bin in the recoil electron energy). This is because
the correlated errors are significant (compared to the statistical errors)
amongst the first four bins and  negligible for the rest. Hence our
our neglect of correlated errors is strictly valid only for these bins.
The result of such an analysis (for $\phi = 0$) is very similar to the
analysis using the entire SNO data set and is shown in Fig.~\ref{fig:lma}
as dashed lines in the second panel. This leads us to believe that our
results are not very sensitive to inclusion of correlations.

Also, for the LMA solution, the $\chi^2$ is smallest for $\phi = 0^\circ$,
increasing with increasing $\phi$; the opposite behaviour is seen for the
SMA solution. In both cases, however, there is a decrease in the allowed
region with increasing $\phi$. This provides a loose bound on $\phi$,
$\phi < 20^\circ$, independently of {\sc chooz} data. This is also seen
from the global minimisation with respect to all three free parameters.
The LMA best fit yields $\phi \sim 0$ with a $1\sigma$ error of about
$23^\circ$ as can be seen from Table~1.

\subsection{The survival probability}

\begin{figure}[htp]
\vskip 9truecm
\includegraphics{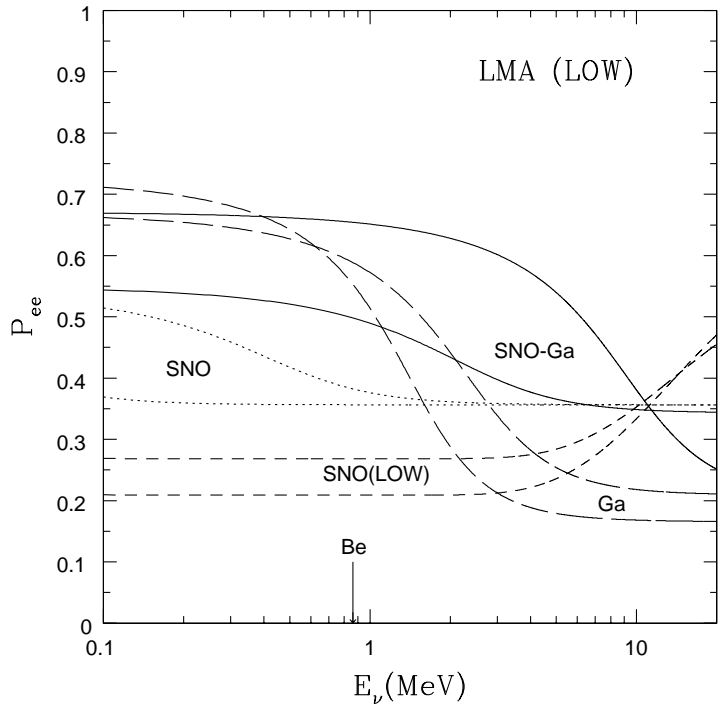}
	  \caption{The electron neutrino survival probability $P_{ee}$
	  as a function of neutrino energy for the best fit solution in
	  the LMA/LOW regions for $\phi = 0^\circ$. Two sets of lines
	  for each solution indicate the $1\sigma$ band in the
	  parameters. Solutions to individual SNO and Ga data are shown
	  as dotted and long-dashed lines; the SNO LOW solution is shown
	  as dashed lines. The solid lines indicate the preferred $P_{ee}$
	  from fits to the combined Gallium and SNO data sets.}
\label{fig:pee_lma} % fig3
\end{figure}

\begin{figure}[htp]
\vskip 9truecm
\includegraphics{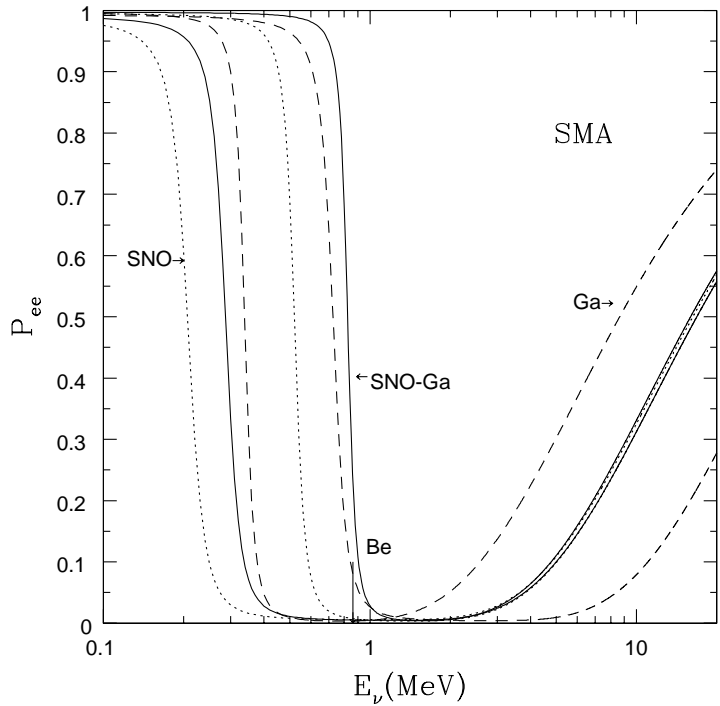}
	  \caption{The electron neutrino survival probability
	  $P_{ee}$ as a function of neutrino energy for the best fit
	  solution in the SMA region for $\phi = 0^\circ$. Two sets of
	  lines for each solution indicate the $1\sigma$ band in the
	  parameters. Solutions to individual SNO and Ga data are shown
	  as dotted and long-dashed lines. The solid lines indicate
	  the preferred $P_{ee}$ from fits to the combined Gallium and
	  SNO data sets.}
\label{fig:pee_sma} % fig4
\end{figure}

The probability shapes corresponding to these preferred solutions are
shown in Figs.~\ref{fig:pee_lma} and \ref{fig:pee_sma}. The two
figures correspond to the so-called large mixing angle (LMA and LOW)
solutions and the small mixing angle (SMA) solutions. While all three
solutions are allowed only by SNO, Ga has a preferred SMA and LMA
solution, whose corresponding $P_{ee}$ is shown in the figures.
The $P_{ee}$ corresponding to the $\chi^2$ minimum from the combined
Ga-SNO analysis is shown as solid lines in the figures.  The two lines
per solution correspond to the $1\sigma$ limits on the parameter values
and indicate the range of $P_{ee}$ allowed.

The energy corresponding to the (second) ${}^7$Be line is indicated on
both the figures for LMA/LOW and SMA solutions.  The LMA and SMA solutions
clearly show that roughly 60\% and 10\% respectively of the total Be flux
will survive and be detected at Borexino, for example. Even beyond the
$1\sigma$ level, a low (less than about 30\%) fraction of ${}^7Be$ will
indicate a clear preference for the SMA solution. Hence a measurement of
the Be flux will be crucial in distinguishing the LMA and SMA solutions
and in narrowing down the allowed parameter space for mixing.

It is seen that the results of the combined analysis are driven mostly
by the SNO data. Hence the errors on the $P_{ee}$ are smaller at larger
energies where SNO data are available. In particular, while still being
well within the range allowed by the Ga SMA solution, the allowed
$P_{ee}$ range from the SNO SMA solution at larger energies is very
restricted as can be seen from Fig.~\ref{fig:pee_sma}.

The fits favoured by Ga data, on the other hand, tend to either
underestimate (Ga-LMA) or allow a very broad band of $P_{ee}$ in the
energy range relevant for SNO. The reason for this is obvious: there is no
energy information in the Ga data. In spite of this, it is able to rule
out the LOW solution for SNO. (On the average, the Ga data corresponds
to lower energy of roughly $\langle E_\nu \rangle \sim 0.3$ MeV. The
observed Ga ratio which is $0.58 \pm 0.07$ cannot therefore be accomodated
by the LOW solution as can be seen from Fig.~\ref{fig:pee_lma}).

\begin{figure}[htpb]
\vskip 18truecm
\includegraphics{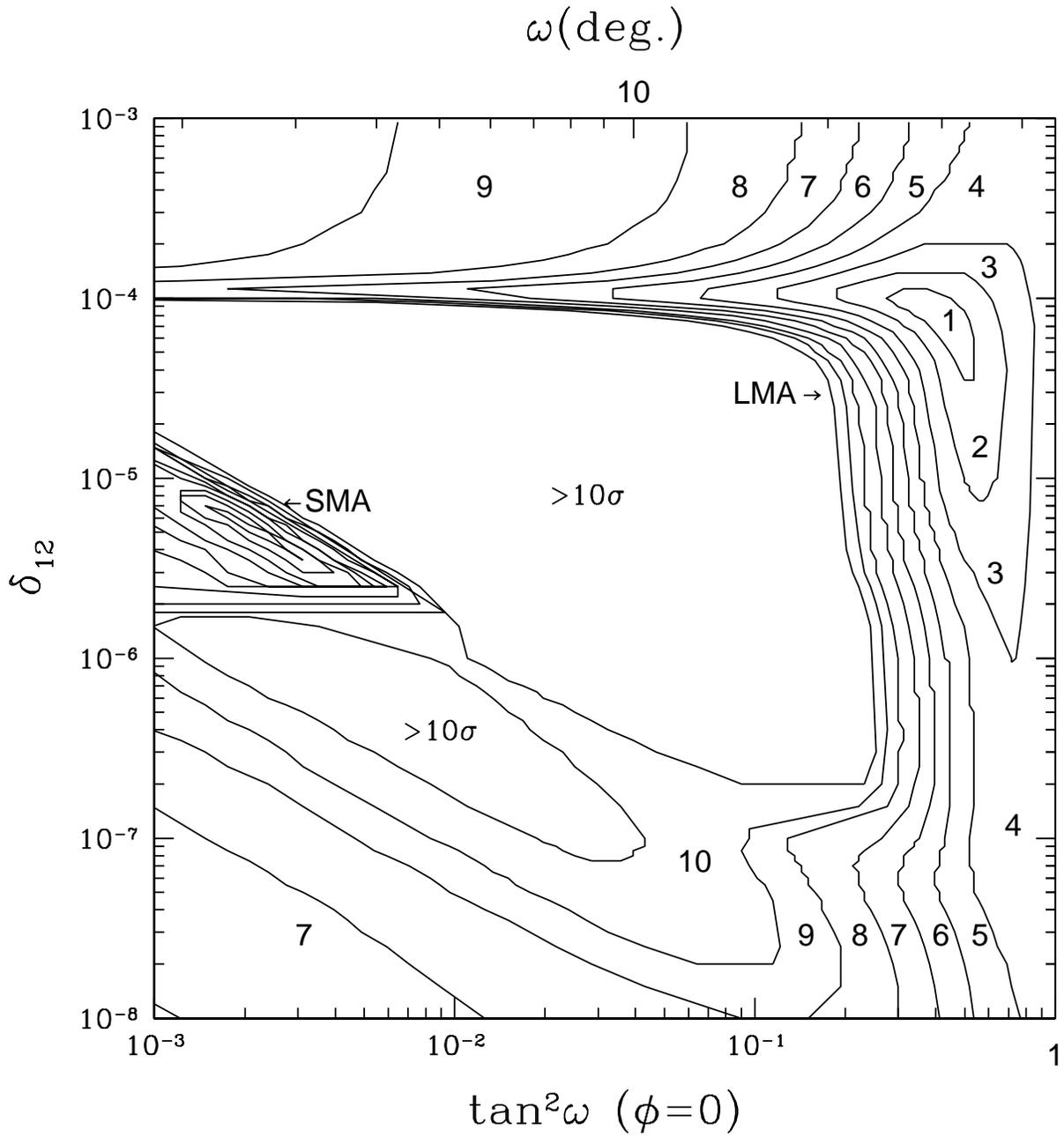}
	  \caption{Likelihood contours in the
	  $\delta_{12}$-$\tan^2\omega$ plane for $\phi = 0^\circ$
	  corresponding to $n\sigma$ deviations, $n = 1, 10$, from the
	  global minimum value of $\chi^2$ for the combined Ga and SNO
	  data fits. The LMA and SMA allowed regions are marked. The
	  central regions of the parameter space are excluded by more
	  than $10\sigma$.}
\label{fig:nlike} % fig5
\end{figure}

Another way of understanding this is to look at the likelihood contours
for the combined Ga-SNO fits over the entire parameter region. We have
shown this in Fig.~\ref{fig:nlike} for $\phi=0$. The likelihood contours
corresponding to $n\sigma$ allowed regions, $n=1,10$, with respect
to the global minimum (see Table~1) are shown in the $\delta_{12}$ vs
$\tan^2\omega$ (or $\omega$) plane. The $1\sigma$ contours show the
two distinct SMA and LMA allowed regions very clearly. It is seen that
the minimum coresponding to SMA is very narrow. The LMA region is larger
and smoothly goes into the so-called ``LOW'' at $4\sigma$. Hence the LOW
solution is allowed only at $4\sigma$. The results are not very
different for $\phi=9^\circ$.

\section{Summary}

We have performed an analysis of the charged current SNO and Gallium
solar neutrino data with a view to understanding the physically allowed
electron neutrino survival probability assuming an oscillation
hypothesis with 3 active neutrino flavours. We have analysed the data
separately as well as done a combined fit. The results can be summarised
as follows.

\begin{figure}[htp]
\vskip 20truecm
\includegraphics{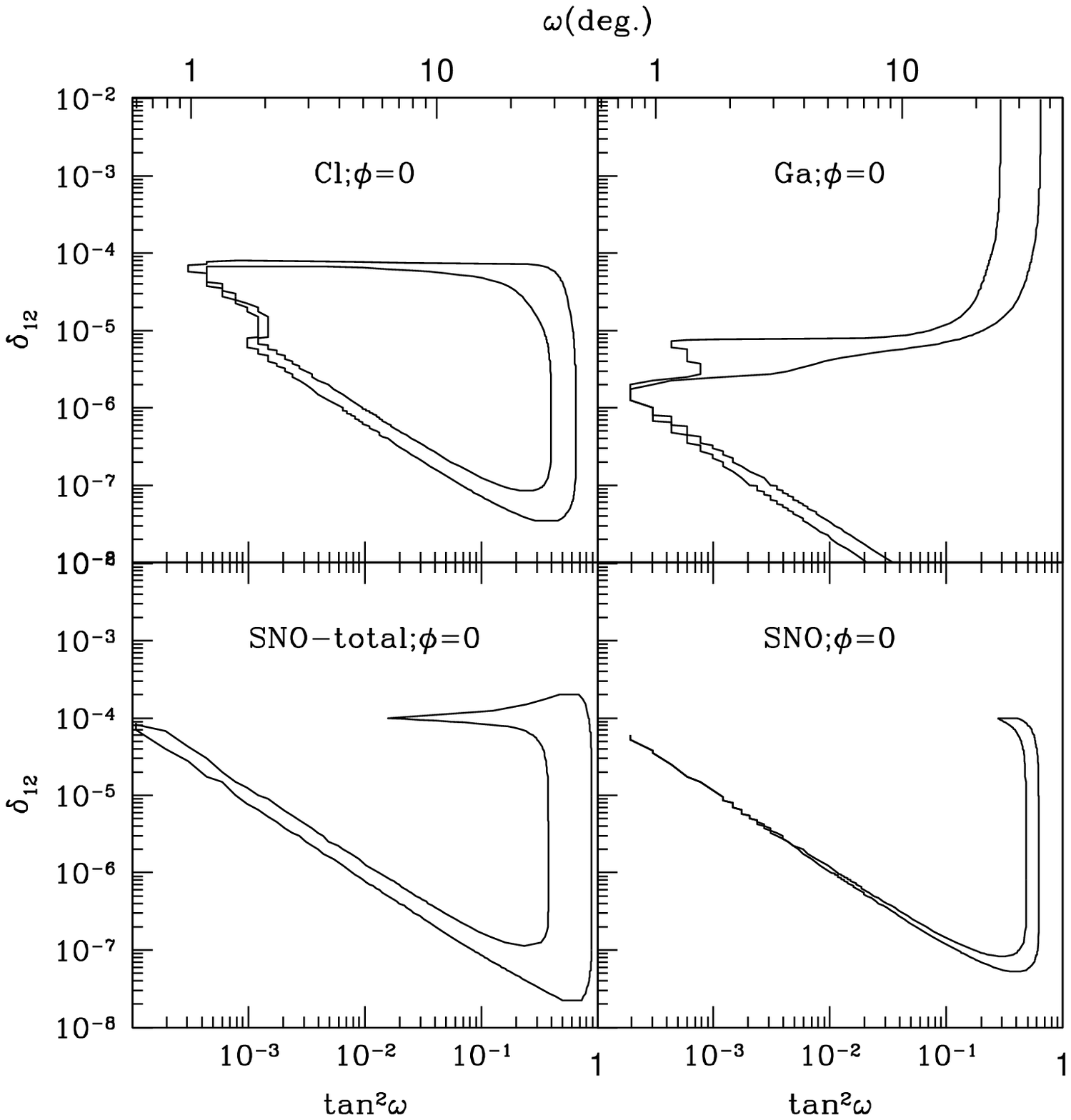}
	  \caption{1$\sigma$ allowed contours at $\phi = 0$ for the
	  individual Chlorine, Gallium and SNO data sets. Results for
	  the SNO total integrated data as well as for the
	  energy-binned data are separately shown in the bottom panels.}
\label{fig:1sigma} % fig1
\end{figure}

\begin{itemize}
\item The data are not very sensitive to the (13) mixing angle, $\phi$.
However even if we do not apply the {\sc chooz} bound \cite{chooz},
$\phi < 9^\circ$, very large $\phi > 20^\circ$ does not seem compatible
with the SNO/Ga data. 

\item The individual Gallium and SNO (as also the Chlorine data)
give rise to very different allowed regions of parameter space. These
regions are contiguous with no distinct LMA and SMA regions visible (see
Fig.~\ref{fig:1sigma}.

\item In particular the so-called 'LOW' solution (small (12) mass-squared
difference, $\delta_{12}$, large (12) mixing angle, $\omega$) is totally
incompatible with the Ga data although it is the preferred solution for
SNO.  A combined analysis of the SNO and Ga data rules out this solution
at $4\sigma$.

\item Distinct small angle and large angle solutions (SMA and LMA)
appear on combining the Ga and SNO data sets. Both solutions are equally
preferred.

\item Finally, the currently allowed SMA and LMA solutions predict very
different survival probability of the (second) ${}^7$Be line spectrum.
An exclusive measurement of the ${}^7$Be flux such as from
Borexino will be able to distinguish whether the small angle solution or
the large angle solution is ultimately the right solution.
\end{itemize}

Due to the severity of the {\sc chooz} constraint \cite{DIMR01}, it is
possible that these results are not significantly altered in the
presence of a fourth (sterile) neutrino flavour.

It may be pointed out that other analyses \cite{analysis} find all
three allowed regions, LMA, SMA and LOW. While the Ga experiments yield
one data point (since it is an energy integrated measurement), the SNO
and Super-Kamiokande experiments correspond to many points because of
energy information. It is possible that the significance of the Ga data
is reduced when doing a combined $\chi^2$ analysis of such a data set.

A combination of the Ga, Cl and SNO (integrated) data can be used to
determine certain ``energy-averaged' values of $P_{ee}$ \cite{BL}. For
example, a constant $P_{ee}$ corresponding to the SNO data at an
average energy $\langle E_\nu \rangle = 10.8$ MeV may be assumed. Then
the SNO data determine the value of $P_{ee}$ at this energy. Assuming
that this value holds for all ${}^8$B neutrinos, we obtain the central
value, $P_{ee}(E_\nu^{\rm B}) = 0.347$ from SNO data.  The Cl data has
contributions from the ${}^8$B flux as well as the ${}^7$Be flux. Using
$P_{ee}(E_\nu^{\rm B})$, we obtain $P_{ee}(E_\nu^{\rm Be}) = 0.29$;
finally, using these values in the Ga data, the average survival
probability in the pp region is obtained to be $P_{ee}(E_\nu^{\rm pp})
= 0.76$. Since $\langle E_\nu^{\rm pp}\rangle = 0.3$ MeV, and $E_\nu^{\rm
Be} = 0.86$ MeV, while $\langle E_\nu^{\rm B} \rangle \sim 9$ MeV,
it therefore appears as though there is a large energy dependence of
$P_{ee}$ in the lower energy $E_\nu < 1$ MeV regime.  However, it is
crucial to realise that this simplistic analysis may not hold when a
detailed energy analysis is considered. For example the LMA solution that
we have obtained corresponds to a flatter $P_{ee}$ around 0.6 in this low
energy region while the SMA corresponds to a steeper one with an almost
vanishing $P_{ee} (E_\nu^{\rm Be}$. In either case the energy averaged
predictions do not hold in detail. A completely unambiguous determination
of $P_{ee}$ in the $E_\nu < 1$ MeV regime must therefore await either
results from Borexino or some new energy-sensitive data in the pp sector.

\acknowledgements{ We thank K. Kubodera, S. Nakamura and T.
Sato for information on the deuterium cross-sections, Jimmy Law for
discussions on the SNO data, and G. Rajasekaran for a critical reading
of the manuscript.}

\newpage
~

\newpage
~

\newpage
~

\newpage
~

\begin{table}
\begin{tabular}{lrll}
$\delta_{12}$   &  $\omega$      &      $\phi$    &  $\chi^2$ \\ \hline
$(8.5 (-5.9, +2.8))\times 10^{-5}$ &  $32.0 (-4.9, +4.7)$ &  0  &   3.1 \\
$(5.9 (-3.1, +2.1))\times 10^{-6}$ &   $2.5 (-0.4, +1.2)$ &  0  &   2.6 \\ \hline

$(1.0 (-0.4, +0.2))\times 10^{-4}$ &  $30.2 (-4.8, +5.6)$ &  9  &   3.3 \\
$(7.2 (-1.6, +0.7))\times 10^{-6}$ &   $2.2 (-0.2, +1.3)$ &  9  &   2.6 \\ \hline

$(1.2 (-0.1, +0.3))\times 10^{-4}$ &  $23.9 (-4.8, +5.0)$ & 20  &   3.9 \\
$(7.3 (-0.4, +0.5))\times 10^{-6}$ &   $2.0 (-0.1, +0.1)$ & 20  &   2.4
\\ \hline
$(8.7 (-6.1, +4.8))\times 10^{-5}$ &  $32.1 (-10.9, +4.9)$ &  $0.003
(-23.0, +23.0)$  &   2.8 \\
\end{tabular}
\caption{Values of $\delta_{12}$ and $\omega$ corresponding to $\chi^2$
minima for the combined fits to SNO and Ga data for different fixed
values of $\phi$. $1\sigma$ errors on the parameter values are also
given. The last row corresponds to a global minimum with respect to all
three parameters, $\delta_{12}$, $\omega$, and $\phi$.}
\end{table}

\begin{references}

\bibitem{sno} SNO Collab., Q.R. Ahmad et al.,  Phys. Rev. Lett. {\bf 87},
071301 (2001); nucl-ex/0106015. The data tables are from the web-site,
http://ewiserver.npl.washington.edu/sno/.

\bibitem{analysis}  For example, see
G.L. Fogli, E. Lisi, D. Montanino, and A. Palazzo, Phys. Rev. {\bf D
64}, 093007 (2001);
J.N. Bahcall, M.C. Gonzalez-Garcia, C. Pena-Garay, JHEP {\bf 0108}, 014
(2001);
A. Bandyopadhyay, S. Choubey, S. Goswami, and K. Kar, Phys. Lett. {\bf B
519}, 83 (2001);
P.I. Krastev, A.Yu. Smirnov, preprint hep-ph/0108177.

\bibitem{sage} SAGE Collab., J.N. Abdurashitov et al., Phys. Rev. Lett.
{\bf 83}, 4686 (1999).

\bibitem{gg} GNO Collab., M. Altmann et al., Phys. Lett. {\bf B 490}, 16
(2000). 

\bibitem{homestake} B.T. Cleveland et al., Astrophys. J. {\bf 496}, 505
(1998).

\bibitem{bp2001} J.N. Bahcall. M.H. Pinsonneault and S. Basu, Astrophys
J. {\bf 555}, 990 (2001). The data tables are taken from the web-site
http://www.sns.ias.edu/~jnb/index.html.

\bibitem{BL}
V. Barger, D. Marfatia, and K. Whisnant, preprint hep-ph/0106207;
V. Berezinsky, M. Lissia, Phys. Lett. {\bf B 521}, 287 (2001).

\bibitem{KuoP} T.K. Kuo and J. Pantaleone, Phys. Rev. {\bf D 35}, 3432
(1987).

\bibitem{bahcallga} J.N. Bahcall, Phys. Rev. {\bf C 56}, 3391 (1997);
hep-ph/9710491.

\bibitem{Kubodera} S. Nakamura, T. Sato, V. Gudkov, and K. Kubodera,
Phys. Rev. {\bf C 63}, 034617 (2001). The data tables are available at
http://nuc003.psc.sc.edu/~kubodera/NU-D-NSGK.

\bibitem{chooz} M. Apollonio et al., {\sc chooz} Collab., Phys. Lett. B {\bf
420}, 397 (1998); Phys. Lett. B {\bf 466}, 415 (1999) (hep-ex/9907037).

\bibitem{DIMR01} Gautam Dutta, D. Indumathi, M.V.N. Murthy, and
G. Rajasekaran, Phys. Rev. {\bf D 64}, 073011 (2001).

\end{references}
\end{document}